\documentclass[12pt,a4paper]{article}

\usepackage{epsfig}

\newcommand{\modii}{\mathrm{I\!I}}
\newcommand{\ZLLi}{c^i_{LL}}
\newcommand{\ZLRi}{c^i_{LR}}
\newcommand{\ZRLi}{c^i_{RL}}

\begin{document}

\vspace*{-3cm}
\begin{flushright}
FISIST/17--2002/CFIF \\
hep-ph/0210360 \\
October 2002
\end{flushright}
\vspace{0.5cm}
\begin{center}
\begin{Large}
{\bf Rare top decays $t \to c \gamma$, $t \to cg$ and CKM unitarity}
\end{Large}

\vspace{0.5cm}
J. A. Aguilar--Saavedra and B. M. Nobre \\
{\it Departamento de F\'{\i}sica and Grupo de F\'{\i}sica de
Part\'{\i}culas (GFP), \\
Instituto Superior T\'ecnico, P-1049-001 Lisboa, Portugal}
\end{center}

\begin{abstract}
Top flavour-changing neutral decays are extremely suppressed within the Standard
Model (SM) by the GIM mechanism, but can reach observable rates in some of its
extensions. We compute the branching ratios for $t \to c \gamma$ and $t \to cg$
in minimal SM extensions where the addition of a vector-like up or down quark
singlet breaks the unitarity of the $3 \times 3$ Cabibbo-Kobayashi-Maskawa
(CKM) matrix. The maximum rates obtained indicate to what extent present
experimental data allow $3 \times 3$ CKM unitarity to be broken in these models,
and are too small to be observed in the near future. As a by-product, we
reproduce the calculation of these branching ratios in the SM, and with an
improved set of parameters we obtain values one order of magnitude smaller than
the ones usually quoted in the literature. We study the CP asymmetries between
the decay rates of the top quark and antiquark, which can be much larger than in
the SM, also as a consequence of the partial breaking of $3 \times 3$ CKM
unitarity.
\end{abstract}

\section{Introduction}

The arrival of top factories, LHC and TESLA, will bring a tremendous improvement
in our knowledge of top quark properties \cite{papiro1,papiro2}. In particular,
the large top samples produced will allow to perform precision studies of top
rare decays. In this field, flavour-changing neutral (FCN) decays $t \to c Z$,
$t \to c \gamma$,
$t \to cg$, deserve special attention. Within the Standard Model (SM) they are
mediated at lowest order in perturbation theory by penguin diagrams with
quarks of charge $Q=-1/3$ inside the loop. Due to the smallness of
down-type quark masses compared to $M_W$, these decays
are very suppressed by
the GIM mechanism, in contrast with processes like
$b \to s \gamma$, with diagrams with a top
quark in the loop. This extra suppression results in decay rates $O(10^{-10})$
or smaller \cite{papiro3}.
On the other hand, in several SM extensions the branching
ratios for FCN top decays can be orders of magnitude larger. For instance, in
two Higgs doublet models $\mathrm{Br}(t \to c Z) \sim 10^{-6}$,
$\mathrm{Br}(t \to c \gamma) \sim 10^{-7}$,
$\mathrm{Br}(t \to c g) \sim 10^{-5}$ can be achieved \cite{papiro4}.
In supersymmetric models with $R$ parity conservation these branching ratios
can reach $\mathrm{Br}(t \to c Z) \sim 10^{-6}$,
$\mathrm{Br}(t \to c \gamma) \sim 10^{-6}$,
$\mathrm{Br}(t \to c g) \sim 10^{-5}$ \cite{papiro5,papiro6}.

Here we are interested in the possible enhancement of these rates in models with
vector-like quark singlets. The addition of quark singlets to the SM
particle content represents the simplest way to break the GIM mechanism
consistently. In these models, the $3 \times 3$ Cabibbo-Kobayashi-Maskawa (CKM)
matrix
is not unitary, and thus FCN couplings to the $Z$
boson appear at tree-level. FCN couplings between light quarks are
experimentally constrained to be very small, but this is not the case for
the top quark. Actually, top FCN vertices can mediate
the decays $t \to u Z$ and $t \to c Z$, giving observable rates in models with
up-type singlets \cite{papiro7}.
The largest branching ratios allowed by present experimental data are
$\mathrm{Br}(t \to u Z) = 7.0 \times 10^{-4}$,
$\mathrm{Br}(t \to c Z) = 6.0 \times 10^{-4}$ \cite{papiro8}, 
much smaller than present direct limits $\mathrm{Br}(t \to uZ),\;
\mathrm{Br}(t \to cZ) \leq 0.08$ \cite{papiro21} but still
observable at LHC
\cite{papiro17a,papiro17b,papiro17c} and TESLA
\cite{papiro18a,papiro18b,papiro18c}.
In this Letter we investigate the enhancement of the branching ratios for the
two other FCN decays, $t \to c \gamma$ and $t \to c g$, in the presence of
either up or down singlets. We find the rates of these processes
allowed by present experimental constraints, and study how the GIM suppression
takes place in these models. For completeness
we also quote without discussion the results for $t \to u \gamma$ and
$t \to u g$, which in the
SM are suppressed by the ratio $|V_{ub}/V_{cb}|^2$ with respect to the former,
but in these SM extensions can have the same magnitude.

\section{Overview of the Lagrangian}

A full discussion of the Lagrangian in the weak eigenstate and mass eigenstate
bases can be found for instance in Refs.~\cite{papiro9a,papiro9b}. Here we only
collect the terms of the Lagrangian in the mass eigenstate basis
relevant for our study. 
We consider the SM extended with $n_u$ up singlets and $n_d$ down singlets,
with $n_u$, $n_d$ arbitrary for the moment. The
charged current Lagrangian is
\begin{equation}
\mathcal{L}_W = -\frac{g}{\sqrt 2} \bar u_L \gamma^\mu V d_L W_\mu^+
+ \mathrm{h.c.} \,,
\label{ec:1}
\end{equation}
with $V$ the generalised CKM matrix, of dimension $(3+n_u) \times (3+n_d)$.
The neutral-current Lagrangian describing the interactions with the $Z$ boson 
is
\begin{equation}
\mathcal{L}_Z = -\frac{g}{2 c_W} \left(
\bar u_L \gamma^\mu X^u u_L - \bar d_L \gamma^\mu X^d d_L 
- 2 s_W^2 J_\mathrm{EM}^\mu \right) Z_\mu \,,
\label{ec:2}
\end{equation}
where $X^u$, $X^d$ are hermitian matrices of dimension
$(3+n_u) \times (3+n_u)$ and $(3+n_d) \times (3+n_d)$, respectively. These
matrices can be related to the CKM matrix by $X^u = V \, V^\dagger$,
$X^d = V^\dagger \, V$. The interactions with the unphysical charged scalars
$\phi^\pm$ are given by
\begin{equation}
\mathcal{L}_\phi = -\frac{g}{\sqrt 2 M_W}  \bar u \left( \mathcal{M}^u 
V P_L - V \mathcal{M}^d P_R \right) d \, \phi^+ + \mathrm{h.c.} \,,
\label{ec:3}
\end{equation}
with $\mathcal{M}^u$ and $\mathcal{M}^d$ the diagonal mass matrices for the up
and down quarks. The terms corresponding to the unphysical neutral scalar $\chi$
are
\begin{eqnarray}
\mathcal{L}_\chi & = & \frac{i\,g}{2 M_W} \left[ 
\bar u \left(  \mathcal{M}^u X^u P_L  - X^u \mathcal{M}^u P_R \right) u
\right. \nonumber \\ 
& & \left. - \bar d \left( \mathcal{M}^d  X^d P_L - X^d \mathcal{M}^d P_R
\right) d \right] \chi \,.
\label{ec:4}
\end{eqnarray}
Finally, the terms describing the interactions with the Higgs boson are
\begin{eqnarray}
\mathcal{L}_H & = & \frac{g}{2 M_W} \left[ 
\bar u \left(  \mathcal{M}^u X^u P_L  + X^u \mathcal{M}^u P_R \right) u
\right. \nonumber \\ 
& & \left. + \bar d \left( \mathcal{M}^d  X^d P_L + X^d \mathcal{M}^d P_R
\right) d \right] H \,.
\label{ec:5}
\end{eqnarray}
In our analysis of $t \to c \gamma$ and $t \to c g$ we discuss the two 
simplest cases: $n_u=1$, $n_d=0$ (which will be called Model I) and $n_u=0$,
$n_d=1$ (Model $\modii$). These two cases correspond
to CKM matrices of dimension $4 \times 3$ and $3 \times 4$, respectively, and in
both models the CKM matrix is a submatrix of a unitary $4 \times 4$
matrix.

\section{Calculation of the decay rates}

\begin{figure}[htb]
\begin{center}
\begin{tabular}{ccc}
\mbox{\epsfig{file=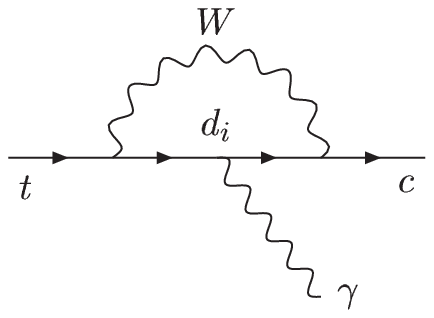,width=4cm,clip=}}  & ~~~ &
\mbox{\epsfig{file=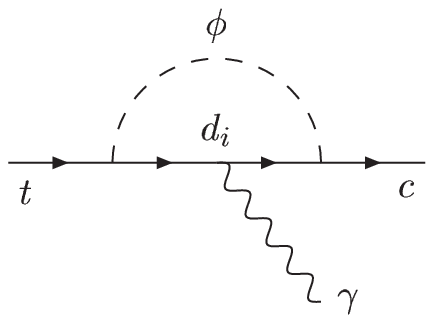,width=4cm,clip=}} \\[-0.2cm]
(1a) & & (1b) \\[0.6cm]
\mbox{\epsfig{file=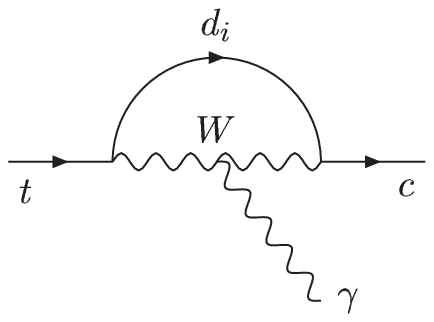,width=4cm,clip=}} & ~~~ &
\mbox{\epsfig{file=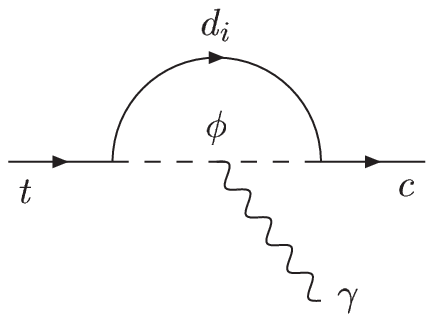,width=4cm,clip=}} \\[-0.2cm]
(2a) & & (2b) \\[0.6cm]
\mbox{\epsfig{file=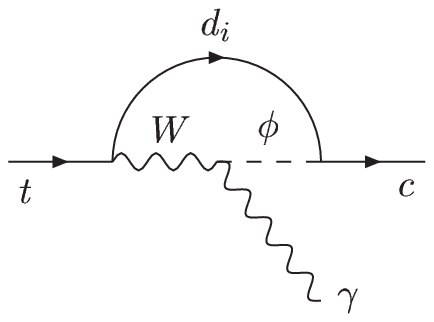,width=4cm,clip=}} & ~~~ &
\mbox{\epsfig{file=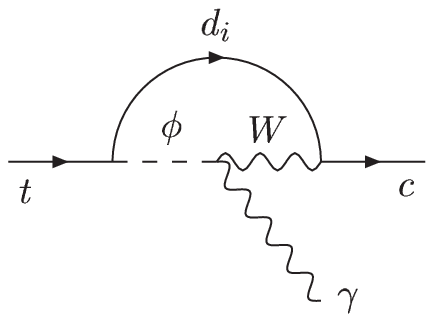,width=4cm,clip=}} \\[-0.2cm]
(2c) & & (2d) \\
\end{tabular}
\caption{Feynman diagrams contributing to the $t \to c \gamma$ decay amplitude
in the SM and Model $\modii$.
\label{fig:1}}
\end{center}
\end{figure}

Using unbroken $\mathrm{SU}(3) \times \mathrm{U}(1)$ gauge invariance and the
facts that final state particles are on-shell and the photon has transverse
polarisation, the transition amplitude for $t \to c \gamma$ can be written
with all generality as
\begin{eqnarray}
\mathcal{M}(t \to c \gamma) & = & \bar u(p_c) \left[ i \sigma^{\mu \nu} q_\nu
\left( A_\gamma + B_\gamma \gamma_5 \right) \right] u(p_t) \epsilon_\mu^*(q)
\,,
\label{ec:6}
\end{eqnarray}
with $p_t$ and $p_c$ the momenta of the top and charm quarks, respectively,
$q=p_t-p_c$ the photon momentum and $\epsilon$ its polarisation vector. This
expression also assumes that the top quark is on-shell, which is an
excellent approximation. A similar structure is valid for $t \to c g$, with form
factors $A_g$ and $B_g$. In order to compute the amplitude the form factors are
written in terms of Passarino-Veltman
integrals \cite{papiro10} using {\tt FORM} \cite{papiro11}. The integrals are
numerically evaluated using {\tt LoopTools} \cite{papiro12}. The Feynman
diagrams relevant for $t \to c \gamma$ in the SM and Model $\modii$ are
depicted in Fig.~\ref{fig:1}. 
In the SM the down-type quarks $d_i$ in the loops
are $d_i = d,s,b$, while in Model $\modii$ there is an extra heavy quark $B$.
The contributions of these diagrams to $A_\gamma$
and $B_\gamma$ in the 't Hooft-Feynman gauge are collected in the Appendix.
The diagrams relevant for $A_g$ and $B_g$ are the analogous to (1a) and
(1b) in Fig.~\ref{fig:1} but replacing the outgoing photon by a gluon.

\begin{figure}[htb]
\begin{center}
\begin{tabular}{ccc}
\mbox{\epsfig{file=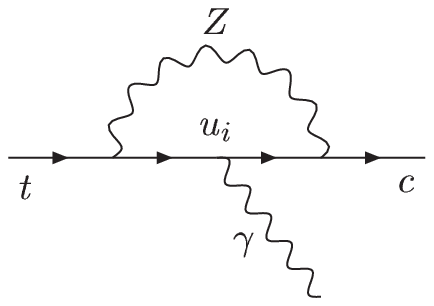,width=4cm,clip=}}  & ~~~ &
\mbox{\epsfig{file=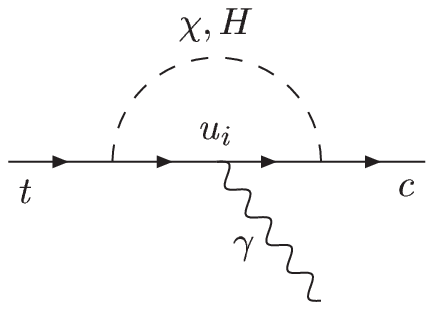,width=4cm,clip=}} \\
(3a) & & (3b,3c)
\end{tabular}
\caption{Additional Feynman diagrams contributing to the $t \to c \gamma$ decay
amplitude in Model I. The diagrams for $t \to c g$ are similar,
replacing the photon by a gluon.
\label{fig:2}}
\end{center}
\end{figure}

In Model I there are extra diagrams with up-type quarks $u_i=u,c,t,T$ in the
loops (see Fig.~\ref{fig:2}).
 These diagrams have one FCN vertex for $u_i=c,t$ and two for $u_i=u,T$,
in which case they are very suppressed. The flavour-diagonal vertices are
modified with respect to the SM value. For instance,
the diagonal couplings of a quark $q=u_i,d_i$ to the $Z$ boson are
\begin{eqnarray}
c_L^q & = & \pm X_{qq} -2 Q_q s_W^2 \,, \nonumber \\
c_R^q & = & -2 Q_q s_W^2 \,,
\label{ec:7}
\end{eqnarray}
as can be seen from Eq.~(\ref{ec:2}),
with the plus (minus) sign for up (down) quarks. The interactions with the
unphysical scalar $\chi$ and the Higgs boson can be read from
Eqs.~(\ref{ec:4},\ref{ec:5}). The contributions of these diagrams to
$A_\gamma$ and $B_\gamma$ can be found in the Appendix.

We perform the computation keeping all quark masses. For external quarks
we use the pole masses $m_t = 174.3$ GeV, $m_c = 1.5$ GeV. For internal
quarks it is more adequate to use $\overline \mathrm{MS}$ masses at a scale
$O(m_t)$, rather than pole masses. This is an important difference due to the
strong dependence on the $b$ quark mass as a consequence of the GIM suppression.
For a pole mass $m_b = 4.7 \pm 0.3$ GeV,
${\overline m}_b(m_t) = 2.74 \pm 0.17$ GeV \cite{papiro13}.

In the limit $m_c=0$ the vector and axial form factors are equal:
$A_\gamma = B_\gamma$, $A_g = B_g$. Since $m_c$ is small,
$A_\gamma \simeq B_\gamma$, $A_g \simeq B_g$ and the effective
couplings are predominantly right-handed.
One important feature is that the form
factors acquire imaginary parts from the contributions with $d$, $s$, $b$
quarks (and $u$, $c$ quarks in the extra diagrams present in Model I).
These imaginary parts are one of the ingredients needed in order to have CP
asymmetries $\Gamma (t \to c \gamma) \neq \Gamma (\bar t \to \bar c \gamma)$,
which will be analysed in detail later. 

From Eq.~(\ref{ec:6}), the partial widths of these processes are
\begin{eqnarray}
\Gamma (t \to c \gamma) & = & \frac{1}{\pi} 
\left[ \frac{m_t^2-m_c^2}{2 m_t} \right]^3
\left( |A_\gamma|^2+|B_\gamma|^2 \right) \,,
\nonumber \\
\Gamma (t \to c g) & = & \frac{C_F}{\pi} 
\left[ \frac{m_t^2-m_c^2}{2 m_t} \right]^3  \left( |A_g|^2+|B_g|^2 \right) \,,
\label{ec:8}
\end{eqnarray}
with $C_F=4/3$ a colour factor. In the SM, as well as in our models, the total
width is dominated by the leading decay mode $t \to b W^+$,
$\Gamma(t \to b W^+) = 1.57 \, |V_{tb}|^2$ for $m_t=174.3$ GeV,
$M_W = 80.39$ GeV. 
The branching ratios are then
\begin{equation}
\mathrm{Br}(t \to c \gamma) = 
\frac{\Gamma (t \to c \gamma)}{\Gamma (t \to b W^+)} ~,~~
\mathrm{Br}(t \to c g) = 
\frac{\Gamma (t \to c g)}{\Gamma (t \to b W^+)} \,.
\label{ec:9}
\end{equation}
We do not use the next-to-leading order
partial width $\Gamma (t \to b W^+) = 1.42 \, |V_{tb}|^2$
for consistency, because our calculation for $t \to c\gamma$, $t \to cg$ is
at leading order.

We have checked that using the set of input parameters of
Ref.~\cite{papiro3} our results agree with the results presented there.
For the calculation of the branching ratios within the SM
we take $|V_{us}| = 0.2224$, $|V_{ub}| = 0.00362$, $|V_{cb}| = 0.0402$. These
values are obtained performing a fit to the six measured CKM matrix elements,
using $3 \times 3$ unitarity. The phase $\delta$ in the standard
parameterisation \cite{papiro14} is $\delta = 1.014$, obtained with a fit
to $\varepsilon$, $\varepsilon'/\varepsilon$, $a_{\psi K_S}$ and $|\delta m_B|$
(see Ref.~\cite{papiro8}). The SM predictions are
\begin{eqnarray}
\mathrm{Br}(t \to c \gamma) & = & (4.6 ~^{+1.2}_{-1.0} \pm 0.4 ~^{+1.6}_{-0.5}
) \times 10^{-14} \,, \nonumber \\[0.2cm]
\mathrm{Br}(t \to c g)      & = & (4.6  ~^{+1.1}_{-0.9} \pm 0.4 ~^{+2.1}_{-0.7}
) \times 10^{-12} \,.
\label{ec:11}
\end{eqnarray}
The first uncertainty comes from the bottom mass, the second from CKM mixing
angles and the third is estimated varying the renormalisation scale between
$M_Z$ (plus sign) and $1.5 \, m_t$ (minus sign). These figures are ten times
smaller than the ones quoted in Ref.~\cite{papiro3}, where the pole mass is used
for the internal $b$ quark ($m_b = 5$ GeV is assumed). The
uncertainty in the top mass does not affect these values, because 
the partial widths of $t \to c \gamma$, $t \to c g$ are proportional to $m_t^3$,
and the partial width of $t \to b W^+$ is approximately given by
\begin{eqnarray}
\Gamma (t \to b W^+) & = & \frac{g^2}{64 \pi} |V_{tb}|^2 \frac{m_t^3}{M_W^2} 
\left[ 1-3 \frac{M_W^4}{m_t^4} + 2 \frac{M_W^6}{m_t^6} \right] \,.
\label{ec:10}
\end{eqnarray}
Hence, the leading dependence on $m_t$ cancels in the ratios and the uncertainty
in $m_t$ hardly affects the numbers quoted in Eqs.~(\ref{ec:11}).
The SM predictions for $t \to u \gamma$ and
$t \to u g$ are
\begin{eqnarray}
\mathrm{Br}(t \to u \gamma) & = & (3.7 ~^{+1.0}_{-0.8} \pm 2.1 ~^{+1.3}_{-0.4}
) \times 10^{-16} \,, \nonumber \\[0.2cm]
\mathrm{Br}(t \to u g)      & = & (3.7  ~^{+0.9}_{-0.8} \pm 2.1 ~^{+1.7}_{-0.5}
) \times 10^{-14} \,,
\label{ec:11b}
\end{eqnarray}
suppressed by a factor $|V_{ub}/V_{cb}|^2 \simeq 8 \times 10^{-3}$ with respect
to top decays to a charm quark. The uncertainties have the same origin as 
in Eqs.~(\ref{ec:11}).

\section{CKM unitarity and GIM suppression}

\begin{figure}[htb]
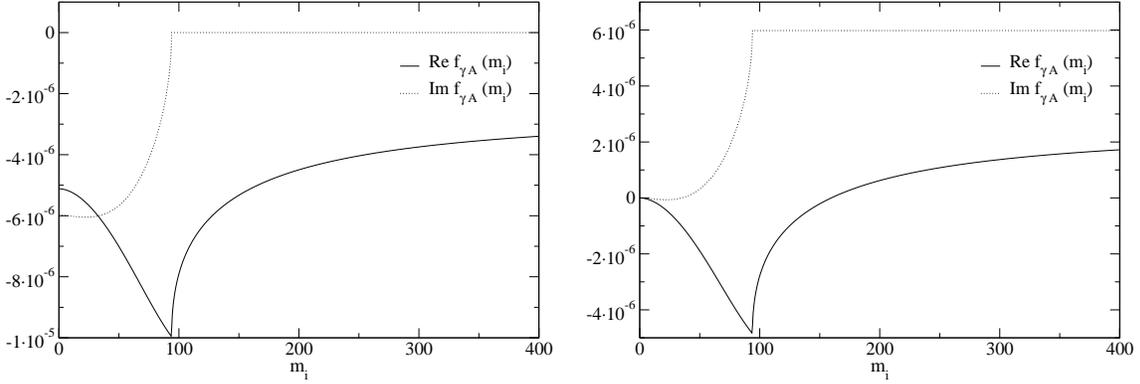

\begin{center}
\begin{tabular}{cc}
\mbox{\epsfig{file=Figs/A.eps,width=7.3cm,clip=}} &
\mbox{\epsfig{file=Figs/Ap.eps,width=7.3cm,clip=}} 
\end{tabular}
\caption{Loop functions $f_{\gamma A}(m_i)$ and $f'_{\gamma A}(m_i)$
for down-type internal quarks (notice the different scales).
\label{fig:3}}
\end{center}
\end{figure}

Let us discuss how the GIM mechanism suppresses these processes within the
SM and how this suppression can be
partially removed with the addition of vector-like singlets.
We only study $t \to c \gamma$, the discussion of $t \to c g$ is formally
identical.
In the SM and Model $\modii$ the form factors for the $\gamma t c$ vertex
can be decomposed as
\begin{eqnarray}
A_\gamma & = & \sum_i f_{\gamma A}(m_i) \lambda_{ct}^i \,, \nonumber \\
B_\gamma & = & \sum_i f_{\gamma B}(m_i) \lambda_{ct}^i \,,
\label{ec:12}
\end{eqnarray}
where $i=1,2,3$ in the SM and $i=1\cdots4$ in Model $\modii$,
$f_{\gamma A}(m_i)$, $f_{\gamma B}(m_i)$ are functions of the internal quark
mass and
$\lambda_{ct}^i = V_{ci} V_{ti}^*$ are CKM factors. We have dropped the bar over
$m_i$, which are understood as $\overline \mathrm{MS}$ masses.
Since $f_{\gamma A}(m_i) \simeq f_{\gamma B}(m_i)$, we only analyse
$A_\gamma$.
The mass dependence of the real and
imaginary parts of $f_{\gamma A}(m_i)$ is shown in Fig.~\ref{fig:3}.

In order to estimate the branching ratio for $t \to c \gamma$ we make the
approximation $f_{\gamma A}(m_d) = f_{\gamma A}(m_s) = f_{\gamma A}(0)$ for the
moment. Then, using
\begin{equation}
\lambda_{ct}^d + \lambda_{ct}^s + \lambda_{ct}^b + \lambda_{ct}^B = 0 \,,
\label{ec:13}
\end{equation}
as implied by the unbroken row unitarity of the $3 \times 4$ CKM matrix $V$, we
have
\begin{eqnarray}
A_\gamma & = & \left[ f_{\gamma A}(m_b)-f_{\gamma A}(0) \right] \lambda_{ct}^b
+ \left[ f_{\gamma A}(m_B)-f_{\gamma A}(0) \right] \lambda_{ct}^B \nonumber \\
& \equiv & f'_{\gamma A}(m_b) \lambda_{ct}^b 
+ f'_{\gamma A}(m_B) \lambda_{ct}^B \,.
\label{ec:14}
\end{eqnarray}
Therefore, the decay amplitude is actually controlled by the shifted function
$f'_{\gamma A}(m_i)$, plotted in Fig.~\ref{fig:3} as well. The parameter
$\lambda_{ct}^B$ measures the orthogonality of the $c$ and $t$ rows of the $3
\times 3$ CKM submatrix $V_{3 \times 3}$ (see Eq.~(\ref{ec:13})), {\em i. e.}
the breaking of the GIM mechanism in this process.
The SM limit is recovered setting, $\lambda_{ct}^B = 0$, so the only
contribution to the form
factor is given by the small function $f'_{\gamma A}(m_b) \simeq -9.1 \times
10^{-9} -4.7 \times 10^{-9} \, i$ multiplied by $\lambda_{ct}^b \simeq 0.04$.
With an extra down quark, there is a new term with a larger function
$f'_{\gamma A}(m_B) \simeq 4.9 \times 10^{-7} + 6.0 \times 10^{-6} \, i$ (for
$m_B = 200$ GeV), which is however suppressed by $\lambda_{ct}^B$.

The parameter $\lambda_{ct}^B$ can be related to the breaking of the
column unitarity of $V_{3 \times 3}$. This is easily understood, because
if the columns of this submatrix are orthogonal, so must be the rows. The
explicit relation can be written using the extension of the Wolfenstein
parameterisation \cite{papiro15a} in Ref.~\cite{papiro15}. Assuming that
$X_{ds},X_{db},X_{sb} \sim \lambda^4$, $1-X_{ss} \sim \lambda^3$
and $1-X_{bb} \sim \lambda^3$, we have
\begin{equation}
-\lambda_{ct}^B = \sum_{i=1}^3 V_{ci} V_{ti}^* = X_{sb}- \lambda X_{db}
+ A \lambda^2 \left( X_{bb}-X_{ss} \right)
-\frac{\lambda^2}{2} X_{sb} + O(\lambda^7) \,.
\label{ec:15}
\end{equation}
This equation shows how the breaking of the orthogonality of the
first three columns of $V$ ``propagates'' to the second and
third rows. The effect of the new quark can be estimated 
with $\lambda \simeq 0.22$, $A \simeq 1$ and the typical values
$X_{db} \sim 10^{-3}$, $X_{sb} \sim 10^{-3}$,
$X_{bb}-X_{ss} \sim 10^{-3}$ \cite{papiro8},
obtaining $\lambda_{ct}^B \simeq X_{sb} \sim 10^{-3}$.
With this value the $B$ contribution is 20 times larger than the $b$ term,
giving $\mathrm{Br}(t \to c \gamma) \sim 10^{-11}$.

In Model I, neglecting for the moment diagrams with two FCN vertices,
$A_\gamma$ can be decomposed as
\begin{equation}
A_\gamma = \sum_{i=1}^3 f_{\gamma A}(m_i) \lambda_{ct}^i + g_{\gamma A} X_{ct}
\,,
\label{ec:16}
\end{equation}
with $g_{\gamma A}$ the sum of the $c$ and $t$ diagram contributions, which is
roughly of the same size as the $f_{\gamma A}$ functions. In this model we have
the relation
\begin{equation}
\lambda_{ct}^d + \lambda_{ct}^s + \lambda_{ct}^b = X_{ct} 
\label{ec:17}
\end{equation}
expressing the non-orthogonality of the second and third rows of the CKM matrix,
of dimension $4 \times 3$ in this case (compare with Eq.~(\ref{ec:13})).
Hence, the form factor is written as
\begin{eqnarray}
A_\gamma & = & \left[ f_{\gamma A}(m_b)-f_{\gamma A}(0) \right] \lambda_{ct}^b
+ \left[ g_{\gamma A}+f_{\gamma A}(0) \right] X_{ct} \nonumber \\
& \equiv & f'_{\gamma A}(m_b) \lambda_{ct}^b 
+ g'_{\gamma A} X_{ct} \,,
\label{ec:18}
\end{eqnarray}
with $g'_{\gamma A} = -4.4 \times 10^{-6} - 4.8 \times 10^{-6} \, i$. In this
model the FCN coupling $X_{ct}$ can be $X_{ct} \sim 0.04$ for $V_{tb} \sim 0.6$
\cite{papiro8},
yielding a branching ratio $\mathrm{Br}(t \to c \gamma) \sim 5 \times 10^{-8}$.

We note that the larger branching ratio achieved in Model I is {\em not} a
consequence of the presence of a tree-level coupling $Ztc$, which appears in
the expressions of the form factors on the same footing as the parameter
$\lambda_{ct}^B$. Moreover, the loop integrals of the new
physics contributions of Model I, 
$g'_{\gamma A} = -4.4 \times 10^{-6} - 4.8 \times 10^{-6} \, i$
and of Model $\modii$,
$f'_{\gamma A}(m_B) \simeq 4.9 \times 10^{-7} + 6.0 \times 10^{-6} \, i$, are
very
similar. The only reason for the larger branching ratio in Model I is that
$X_{ct} \gg \lambda_{ct}^B$, that is, 
unitarity of $V_{3 \times 3}$ can be broken to a lesser extent in Model $\modii$
due to the strong requirements on FCN couplings between light quarks.
Additionally, $V_{tb}$ can be smaller in Model I, and the total top width is
reduced.

We are also interested in the CP asymmetry
\begin{eqnarray}
a_\gamma & = & \frac{\Gamma (t \to c \gamma) - \Gamma (\bar t \to \bar c
\gamma)}{\Gamma (t \to c \gamma) + \Gamma (\bar t \to \bar c \gamma)} \,.
\label{ec:19}
\end{eqnarray}
This interest is mainly academic, because if the branching ratios are
unobservable, even less are the asymmetries. However, the latter show how 
large CP asymmetries at high energy are possible in these SM extensions. Here we
analyse in detail $a_\gamma$ in Model $\modii$,
the results for Model I are similar but more involved.
The form factors for $\bar t \to \bar c \gamma$ are
\begin{eqnarray}
\bar A_\gamma & = & f'_{\gamma A}(m_b) \lambda_{ct}^{b*}
+ f'_{\gamma A}(m_B) \, \lambda_{ct}^{B*} \,,
\label{ec:20}
\end{eqnarray}
and an analogous expression for $\bar B_\gamma$. The asymmetry can be written as
$a_\gamma = N_\gamma/D_\gamma$, with
\begin{eqnarray}
N_\gamma & = & -2 \, \mathrm{Im} \left[ f'_{\gamma A}(m_b) 
 {f'}^*_{\gamma A}(m_B) + f'_{\gamma B}(m_b) {f'}^*_{\gamma B}(m_B) \right]
\mathrm{Im} \left[ \lambda_{ct}^b \lambda_{ct}^{B*}\right] \nonumber \\
D_\gamma & = & \left[ |f'_{\gamma A}(m_b)|^2 + |f'_{\gamma B}(m_b)|^2 \right]
  |\lambda_{ct}^b|^2
+ \left[ |f'_{\gamma A}(m_B)|^2 + |f'_{\gamma B}(m_B)|^2 \right]
  |\lambda_{ct}^B|^2 \nonumber \\
& & + 2 \, \mathrm{Re} \left[ f'_{\gamma A}(m_b) {f'}^*_{\gamma A}(m_B)
+ f'_{\gamma B}(m_b) {f'}^*_{\gamma B}(m_B) \right]
\mathrm{Re} \left[ \lambda_{ct}^b \lambda_{ct}^{B*}\right] \,.
\label{ec:21}
\end{eqnarray}
A few comments are in order:
\begin{enumerate}
\item The CP asymmetry $a_\gamma$ is proportional to the imaginary part of the
rephasing-invariant quartet $\lambda_{ct}^b \lambda_{ct}^{B*} 
= V_{cb} V_{cB}^* V_{tb}^* V_{tB} $. This is expected from general grounds.
In fact, it can be shown that in a model with an extra down singlet {\em all\/}
CP violating observables at high energy (that is, when $m_{u,d,s} \sim 0$
compared to the scale of energy involved) must be
proportional to $\mathrm{Im} \, V_{cb} V_{cB}^* V_{tb}^* V_{tB}$,
$\mathrm{Im} \, V_{cb} V_{cB}^* X_{bB} $,
$\mathrm{Im} \, V_{tb} V_{tB}^* X_{bB} $, or a combination of them
\cite{papiro16}.

\item The SM limit is recovered setting $\lambda_{ct}^B = 0$, obtaining a
vanishing CP asymmetry. It is well known that CP asymmetries at high energy
are very small in the SM \cite{papiro22}, due to: ({\em i\/}) the smallness
of $m_u$, $m_d$ and $m_s$,
what leads to a more efficient GIM cancellation; ({\em ii\/})
the small mixing between the top and the first two generations.
When all the quark masses are kept in the computation a
non-vanishing but a negligible asymmetry $a_\gamma \sim -6 \times 10^{-6}$
is obtained. For the gluon case, the asymmetry $a_g \sim -5 \times 10^{-6}$
is also extremely small.

\item Since CPT invariance requires that the total width of the top and the
antitop are equal, the different partial widths $\Gamma(t \to c \gamma) \neq
\Gamma(\bar t \to \bar c \gamma)$ must be compensated in other channel.
In this case, the compensating decay channels are the SM leading modes
$t \to bW$, $t \to sW$, $t \to dW$ and their conjugate processes
\cite{papiro22}.

\item The large phases in the functions $f'_{\gamma A}$, $f'_{\gamma B}$ allow
to obtain relatively large CP asymmetries, provided
$\mathrm{Im} \, \lambda_{ct}^b \lambda_{ct}^{B*}$ is sizeable.
\end{enumerate}

\section{Results}

We explore the parameter space of Models I and $\modii$ to find the maximum
values of $\mathrm{Br}(t \to c \gamma)$ and $\mathrm{Br}(t \to c g)$
allowed by present experimental measurements. The constraints on these models
come from
precision electroweak data, $K$ and $B$ physics and atomic parity violation
(the details of the analysis can be found in Ref.~\cite{papiro8}). We
take all the quark masses into account, and require
that the mass of the new quark is larger than 200 GeV to satisfy the
limits from direct searches. In Model I,
assuming that the new quark has a mass $m_T = 200$ GeV, we find the maximum
rates
\begin{eqnarray}
\mathrm{Br}(t \to c \gamma) & = & 4.5 \times 10^{-8} \,, \nonumber \\
\mathrm{Br}(t \to c g) & = & 8.9 \times 10^{-7} \,,
\label{ec:22}
\end{eqnarray}
corresponding to $|X_{ct}| = 0.037$, $|V_{tb}| = 0.58$.
(The branching ratios scale with $|X_{ct}|^2$ approximately.)
For larger $m_T$, the allowed values of $|X_{ct}|$
are smaller \cite{papiro8}, and these branching ratios decrease.
The CP asymmetries corresponding to the figures in Eqs.~(\ref{ec:22}) are
negligible,
\begin{eqnarray}
a_\gamma & = &  -0.0006 \,, \nonumber \\
a_g      & = &  -0.002 \,,
\label{ec:23}
\end{eqnarray}
because the rates are dominated by the $X_{ct}$ term. The asymmetries can
have values in the range
$-0.5 \leq a_\gamma \leq 0.4$, $-0.9 \leq a_g \leq 0.6$, but only reach the
boundaries of these intervals for branching ratios much smaller than those in
Eqs.~(\ref{ec:22}). The results for decays to up quarks are a little larger,
\begin{eqnarray}
\mathrm{Br}(t \to u \gamma) & = & 4.6 \times 10^{-8} \,, \nonumber \\
\mathrm{Br}(t \to u g) & = & 9.2 \times 10^{-7} \,.
\label{ec:22b}
\end{eqnarray}
In Model $\modii$, assuming that the mass of the new quark is $m_B = 200$ GeV,
we have
\begin{eqnarray}
\mathrm{Br}(t \to c \gamma) & = & 4.5 \times 10^{-12} \,, \nonumber \\
\mathrm{Br}(t \to c g) & = & 6.6 \times 10^{-11} \,.
\label{ec:24}
\end{eqnarray}
These numbers are almost insensitive to the mass of the new $Q=-1/3$ quark for
$m_B \geq 200$ GeV, as can be seen from
Fig.~\ref{fig:1}, and show a small increase with $m_B$. The corresponding
asymmetries are
\begin{eqnarray}
a_\gamma & = &  -0.05 \,, \nonumber \\
a_g & = &  0.56 \,.
\label{ec:25}
\end{eqnarray}
In general, the CP asymmetries take values in the intervals
$-1 \leq a_\gamma \leq 1$, $-1 \leq a_g \leq 1$. The
decay rates to up quarks are larger,
\begin{eqnarray}
\mathrm{Br}(t \to u \gamma) & = & 7.4 \times 10^{-12} \,, \nonumber \\
\mathrm{Br}(t \to u g) & = & 9.5 \times 10^{-11} \,.
\label{ec:24b}
\end{eqnarray}

The branching ratios in Eqs.~(\ref{ec:22}--\ref{ec:24b}) are too small to be
measurable in the near future. The estimated $3\sigma$ sensitivities of LHC to
these decays are $\mathrm{Br}(t \to c \gamma) = 1.2 \times 10^{-5}$
\cite{papiro19}, 
$\mathrm{Br}(t \to u \gamma) = 3.0 \times 10^{-6}$ \cite{papiro17c},
$\mathrm{Br}(t \to c g) = 2.7 \times 10^{-5}$ and
$\mathrm{Br}(t \to u g) = 4.1 \times 10^{-6}$ \cite{papiro20},
with an integrated luminosity of 100 fb$^{-1}$. The TESLA sensitivity to $t \to
c \gamma$ is better but not enough,
$\mathrm{Br}(t \to c \gamma) = 3.6 \times 10^{-6}$ \cite{papiro18b} with a
centre of mass energy of 800 GeV and a luminosity of 500 fb$^{-1}$.
Hence, we observe that in models with up-type singlets the rates for $t \to qZ$,
$q=u,c$
can be observable but not the rates for $t \to q \gamma$ and $t \to q g$, which
are four and three orders of magnitude smaller, respectively. This fact
contrasts with the results for two Higgs doublet models or supersymmetric
extensions of the SM, where the branching ratios for $t \to cZ$ and
$t \to c\gamma$ are similar, and the branching ratio for $t \to cg$ is one order
of magnitude larger. This difference would allow for a consistency check of the
models, should a positive signal of top FCN decays be discovered.

\vspace{1cm}
\noindent
{\Large \bf Acknowledgements}

\vspace{0.4cm} \noindent
J.A.A.S. thanks F. del \'Aguila and M. P\'erez-Victoria for comments and
discussions.
B.M.N. thanks D.F. Carvalho for useful discussions. We thank F. del \'Aguila
and G.C. Branco for reading of the manuscript. This work has been
supported by the European Community's Human Potential Programme under contract
HTRN--CT--2000--00149 Physics at Colliders and by FCT through project
CERN/FIS/43793/2001. The work of B.M.N. has been supported by FCT under the
grant SFRH/BD/995/2000.

\appendix
\section{Form factors for $t \to c \gamma$ and $t \to c g$}
The contributions to $A_\gamma$ and $B_\gamma$ of diagrams (1a) and (1b) with
an internal quark $d_i$ are
\begin{eqnarray}
A_{\gamma,1\mathrm{a}} & = & - \frac{Q_i g^2 e}{2} \frac{1}{16 \pi^2}
V_{2i} V_{3i}^* \left\{
(m_t+m_c) \, C_0 + (2 m_t+m_c) \, C_1 + (m_t+2 m_c) \, C_2 
\right. \nonumber \\[0.1cm]
& & \left.
+ m_t C_{11}  + (m_t+m_c) \, C_{12} + m_c C_{22}
\right\} \,, \label{ec:ap1} \\
B_{\gamma,1\mathrm{a}} & = & -\frac{Q_i g^2 e}{2} \frac{1}{16 \pi^2}
V_{2i} V_{3i}^* \left\{
(m_t-m_c) \, C_0 + (2 m_t-m_c) \, C_1 + (m_t-2 m_c) \, C_2 
\right. \nonumber \\[0.1cm]
& & \left.
+ m_t C_{11} + (m_t- m_c) \, C_{12} - m_c C_{22}
\right\} \,, \label{ec:ap2} \\
A_{\gamma,1\mathrm{b}} & = & -\frac{Q_i g^2 e}{4 M_W^2} \frac{1}{16 \pi^2}
V_{2i} V_{3i}^* \left\{
m_c (m_t^2 - {\overline m}_{d_i}^2) \, C_1
+ m_t (m_c^2 - {\overline m}_{d_i}^2) \, C_2
\right. \nonumber \\
& & + (m_t {\overline m}_{d_i}^2 + m_c m_t^2) \, C_{11}
+ [m_t (m_c^2 + {\overline m}_{d_i}^2)
+  m_c (m_t^2 + {\overline m}_{d_i}^2) ]\, C_{12}
\nonumber \\[0.1cm]
& & \left.
+ (m_t m_c^2 + m_c {\overline m}_{d_i}^2) \, C_{22}
\right\} \,, \label{ec:ap3} \\
B_{\gamma,1\mathrm{b}} & = & -\frac{Q_i g^2 e}{4 M_W^2} \frac{1}{16 \pi^2}
V_{2i} V_{3i}^* \left\{
-m_c (m_t^2 - {\overline m}_{d_i}^2) \, C_1
+ m_t (m_c^2 - {\overline m}_{d_i}^2) \, C_2 
\right. \nonumber \\
& & + (m_t {\overline m}_{d_i}^2 - m_c m_t^2) \, C_{11}
+ [m_t (m_c^2 + {\overline m}_{d_i}^2) 
- m_c (m_t^2 + {\overline m}_{d_i}^2) ]\, C_{12}
\nonumber \\[0.1cm]
& & \left.
+ (m_t m_c^2 - m_c {\overline m}_{d_i}^2) \, C_{22}
\right\} \label{ec:ap4} \,.
\end{eqnarray}
The $C$'s are functions of the external and internal masses,
$C(m_t^2,0,m_c^2,M_W^2,{\overline m}_{d_i}^2,{\overline m}_{d_i}^2)$
in the notation of Ref.~\cite{papiro12}.
For $t \to c g$ the contributions to the form factors $A_g$ and $B_g$ can be
obtained replacing $e$ by $g_s$ and setting $Q_i=1$ in
Eqs.~(\ref{ec:ap1}--\ref{ec:ap4}). The terms from diagrams (2a)--(2d) are
\begin{eqnarray}
A_{\gamma,2\mathrm{a}} & = & \frac{g^2 e}{4} \frac{1}{16 \pi^2}
V_{2i} V_{3i}^* \left\{
(2 m_t + m_c) \, C_0 + (4 m_t + m_c) \, C_1 + (m_t - m_c) \, C_2
\right. \nonumber \\[0.1cm]
& & \left.
+ 2 m_t C_{11} + 2 (m_t - m_c) \, C_{12}
\right\} \,, \\
B_{\gamma,2\mathrm{a}} & = & \frac{g^2 e}{4} \frac{1}{16 \pi^2}
V_{2i} V_{3i}^* \left\{
(2 m_t - m_c) \, C_0 + (4 m_t - m_c) \, C_1 + (m_t + m_c) \, C_2
\right. \nonumber \\[0.1cm]
& & \left.
+ 2 m_t C_{11}  + 2 (m_t + m_c) \, C_{12}
\right\} \,, \\
A_{\gamma,2\mathrm{b}} & = & \frac{g^2 e}{4 M_W^2} \frac{1}{16 \pi^2}
V_{2i} V_{3i}^* \left\{
m_c (m_t^2 - {\overline m}_{d_i}^2) \, C_1
+ (m_t {\overline m}_{d_i}^2 + m_c m_t^2) \, C_{11} 
\right. \nonumber \\[0.1cm]
& & \left.
+ [m_t ({\overline m}_{d_i}^2 - m_c^2)
+ m_c (m_t^2 - {\overline m}_{d_i}^2) ] \, C_{12}
\right\} \,, \\
B_{\gamma,2\mathrm{b}} & = & \frac{g^2 e}{4 M_W^2} \frac{1}{16 \pi^2}
V_{2i} V_{3i}^* \left\{
- m_c (m_t^2 - {\overline m}_{d_i}^2) \, C_1
+ (m_t {\overline m}_{d_i}^2 - m_c m_t^2) \, C_{11}
\right. \nonumber \\[0.1cm]
& & \left.
+ [m_t ({\overline m}_{d_i}^2 - m_c^2)
- m_c (m_t^2 - {\overline m}_{d_i}^2) ] \, C_{12}
\right\} \,, \\
A_{\gamma,2\mathrm{c}} & = & \frac{g^2 e}{4} \frac{1}{16 \pi^2}
V_{2i} V_{3i}^* \left\{
m_c C_0 + m_c C_1 + m_c C_2
\right\} \,, \\
B_{\gamma,2\mathrm{c}} & = & \frac{g^2 e}{4} \frac{1}{16 \pi^2}
V_{2i} V_{3i}^* \left\{
- m_c C_0 - m_c C_1 - m_c C_2
\right\} \,, \\
A_{\gamma,2\mathrm{d}} & = & \frac{g^2 e}{4} \frac{1}{16 \pi^2}
V_{2i} V_{3i}^* \left\{
- m_t C_2
\right\} \,, \\
B_{\gamma,2\mathrm{d}} & = & \frac{g^2 e}{4} \frac{1}{16 \pi^2}
V_{2i} V_{3i}^* \left\{
- m_t C_2
\right\} \,.
\end{eqnarray}
Here the arguments of the functions are 
$C(m_t^2,m_c^2,0,M_W^2,{\overline m}_{d_i}^2,M_W^2)$. In Model I the
contributions from diagrams (3a) and (3b) with internal quarks $u_i$ are
\begin{eqnarray}
A_{\gamma,3\mathrm{a}} & = & \frac{Q_i g^2 e}{4 c_W^2} \frac{1}{16 \pi^2}
\left\{ \ZLLi \left[
- (m_t + m_c) \, C_0 - (2 m_t + m_c) \, C_1 - (m_t + 2 m_c) \, C_2
\right. \right. \nonumber \\[0.1cm]
& & \left. \left.
- m_t C_{11} - (m_t + m_c) \, C_{12} - m_c C_{22}
\right] 
+ (\ZLRi + \ZRLi) \, 2 {\overline m}_{u_i} \left[ C_0 + C_1 + C_2
\right] \right\} \,,
\nonumber \\
B_{\gamma,3\mathrm{a}} & = & \frac{Q_i g^2 e}{4 c_W^2} \frac{1}{16 \pi^2}
\left\{ \ZLLi \left[
- (m_t - m_c) \, C_0 - (2 m_t - m_c) \, C_1 - (m_t - 2 m_c) \, C_2 
\right. \right. \nonumber \\[0.1cm]
& & \left. \left.
- m_t C_{11} - (m_t - m_c) \, C_{12} + m_c C_{22} \right]
+ (\ZLRi - \ZRLi) \, 2 {\overline m}_{u_i} \left[  C_0 + C_1 + C_2 \right]
\right\} \,,
\nonumber \\
A_{\gamma,3\mathrm{b}} & = & \frac{Q_i g^2 e}{8 M_W^2} \frac{1}{16 \pi^2}
X^u_{2i} X^u_{i3} \left\{
- m_c (m_t^2 - {\overline m}_{u_i}^2) \, C_1 
+ m_t ({\overline m}_{u_i}^2 - m_c^2) \, C_2
\right. \nonumber \\[0.1cm]
& &
- (m_t {\overline m}_{u_i}^2 + m_c m_t^2) \, C_{11}
- \left[ m_t (m_c^2 + {\overline m}_{u_i}^2)
+ m_c (m_t^2  + {\overline m}_{u_i}^2) \right] \, C_{12} 
\nonumber \\[0.1cm]
& & \left.
- (m_t m_c^2 + m_c {\overline m}_{u_i}^2) \, C_{22}
\right\} \,, \\
B_{\gamma,3\mathrm{b}} & = & \frac{Q_i g^2 e}{8 M_W^2} \frac{1}{16 \pi^2}
X^u_{2i} X^u_{i3} \left\{
+ m_c (m_t^2 - {\overline m}_{u_i}^2) \, C_1 
+ m_t ({\overline m}_{u_i}^2 - m_c^2) \, C_2
\right. \nonumber \\[0.1cm]
& &
- (m_t {\overline m}_{u_i}^2 - m_c m_t^2) \, C_{11}
- \left[ m_t (m_c^2 + {\overline m}_{u_i}^2)
- m_c (m_t^2 + {\overline m}_{u_i}^2) \right] \, C_{12} 
\nonumber \\[0.1cm]
& & \left.
- (m_t m_c^2 - m_c {\overline m}_{u_i}^2) \, C_{22}
\right\} \,.
\end{eqnarray}
The constants $\ZLLi$, $\ZLRi$ and $\ZRLi$ are products of left-handed and
right-handed couplings between $(c,i)$ and $(i,t)$,
\begin{eqnarray}
\ZLLi & = & \left( X^u_{ci} - \delta_{ci} \frac{4}{3} s_W^2 \right)
\left( X^u_{it} - \delta_{it} \frac{4}{3} s_W^2 \right) \,, \nonumber \\
\ZLRi & = & \left( X^u_{ci} - \delta_{ci} \frac{4}{3} s_W^2 \right)
\left( - \delta_{it} \frac{4}{3} s_W^2 \right)\,, \nonumber \\
\ZRLi & = & \left( - \delta_{ci} \frac{4}{3} s_W^2 \right)
\left( X^u_{it} - \delta_{it} \frac{4}{3} s_W^2 \right) \,,
\end{eqnarray}
with $\delta_{ij} = 1$ if $i=j$ and zero otherwise.
For these diagrams the arguments of the functions are
$C(m_t^2,0,m_c^2,M_Z^2,{\overline m}_{u_i}^2,{\overline m}_{u_i}^2)$. 
Finally, the Higgs
contribution in diagram (3c) reads
\begin{eqnarray}
A_{\gamma,3\mathrm{c}} & = & -\frac{Q_i g^2 e}{8 M_W^2} \frac{1}{16 \pi^2}
X^u_{2i} X^u_{i3} \left\{
 \left[ 2 m_t {\overline m}_{u_i}^2 + m_c (m_t^2 + {\overline m}_{u_i}^2)
\right] \, C_1
\right. \nonumber \\[0.1cm]
& &
+ \left[ m_t (m_c^2 + {\overline m}_{u_i}^2) + 2 m_c {\overline m}_{u_i}^2
\right] \, C_2 + (m_t {\overline m}_{u_i}^2 +  m_c m_t^2) \, C_{11} 
\nonumber \\[0.1cm]
& & \left.
+ \left[ m_t (m_c^2 + {\overline m}_{u_i}^2) 
+ m_c (m_t^2 + {\overline m}_{u_i}^2) \right] \, C_{12}
+ (m_t m_c^2 + m_c {\overline m}_{u_i}^2) \, C_{22}
\right\} \,, \\
B_{\gamma,3\mathrm{c}} & = & -\frac{Q_i g^2 e}{8 M_W^2} \frac{1}{16 \pi^2}
X^u_{2i} X^u_{i3} \left\{
\left[ 2 m_t {\overline m}_{u_i}^2 - m_c (m_t^2 + {\overline m}_{u_i}^2)
\right]\, C_1
\right. \nonumber \\[0.1cm]
& &
+ \left[ m_t (m_c^2 + {\overline m}_{u_i}^2) - 2 m_c {\overline m}_{u_i}^2
\right] \, C_2 + (m_t {\overline m}_{u_i}^2 - m_c m_t^2) \, C_{11}
\nonumber \\[0.1cm]
& & \left.
+ \left[ m_t (m_c^2 + {\overline m}_{u_i}^2 )
- m_c (m_t^2 + {\overline m}_{u_i}^2) \right] \, C_{12}
+ (m_t m_c^2 - m_c {\overline m}_{u_i}^2) \, C_{22}
\right\} \,,
\end{eqnarray}
where the functions are
$C(m_t^2,0,m_c^2,M_H^2,{\overline m}_{u_i}^2,{\overline m}_{u_i}^2)$. We take
$M_H = 115$ GeV. For $t \to c g$ the extra contributions in Model I can be
obtained replacing $e$ by $g_s$ and setting $Q_i=1$.


\begin{thebibliography}{99}

\bibitem{papiro1}
M. Beneke {\em et al.}, hep-ph/0003033

\bibitem{papiro2}
J. A. Aguilar-Saavedra {\em et al.} [ECFA/DESY LC Physics Working Group
Collaboration], hep-ph/0106315

\bibitem{papiro3}
G. Eilam, J. L. Hewett and A. Soni, Phys. Rev. {\bf D44} (1991) 1473
[Erratum-ibid. {\bf D59} (1999) 039901];
J. L. Hewett, T. Takeuchi and S. Thomas, in ``Electroweak Symmetry Breaking and
Beyond the Standard Model'', ed. T. Barklow {\em et al.}, World Scientific 1996

\bibitem{papiro4}
D. Atwood, L. Reina and A. Soni, Phys. Rev. {\bf D55} (1997) 3156

\bibitem{papiro5}
G. M. de Divitiis, R. Petronzio and L. Silvestrini, Nucl. Phys. {\bf B504}
(1997) 45

\bibitem{papiro6}
J. L. Lopez, D. V. Nanopoulos and R. Rangarajan, Phys. Rev. {\bf D56} (1997)
3100

\bibitem{papiro7}
F. del Aguila, J. A. Aguilar-Saavedra and R. Miquel, Phys. Rev. Lett. {\bf 82}
(1999) 1628

\bibitem{papiro8}
J. A. Aguilar-Saavedra, hep-ph/0210112, Phys. Rev. {\bf D} (in press)

\bibitem{papiro21}
G. Abbiendi {\em et al.}  [OPAL Collaboration], Phys. Lett. {\bf B521}
(2001) 181

\bibitem{papiro17a}
T. Han, R. D. Peccei and X. Zhang, Nucl. Phys. {\bf B454} (1995) 527 
\bibitem{papiro17b}
F. del Aguila, J. A. Aguilar-Saavedra and Ll. Ametller, Phys. Lett. {\bf B462}
(1999) 310 
\bibitem{papiro17c}
F. del Aguila and J. A. Aguilar-Saavedra, Nucl. Phys. {\bf B576} (2000) 56 

\bibitem{papiro18a}
T. Han and J. L. Hewett, Phys. Rev. {\bf D60} (1999) 074015 
\bibitem{papiro18b}
J. A. Aguilar-Saavedra, Phys. Lett. {\bf B502} (2001) 115 
\bibitem{papiro18c}
J.-j. Cao, Z.-h. Xiong and J. M. Yang, hep-ph/0208035

\bibitem{papiro9a}
F. del Aguila and M. J. Bowick, Nucl. Phys. {\bf B224} (1983) 107

\bibitem{papiro9b}
G. C. Branco and L. Lavoura, Nucl. Phys. {\bf B278} (1986) 738

\bibitem{papiro10}
G. Passarino and M. J. Veltman, Nucl. Phys. {\bf B160} (1979) 151

\bibitem{papiro11}
J. A. Vermaseren, math-ph/0010025

\bibitem{papiro12}
T. Hahn and M. P\'erez-Victoria, Comput. Phys. Commun. {\bf 118} (1999) 153

\bibitem{papiro13}
H. Fusaoka and Y. Koide, Phys. Rev. {\bf D57} (1998) 3986 

\bibitem{papiro14}
K. Hagiwara {\em et al.}, Particle Data Group, Phys. Rev. {\bf D66}
(2002) 010001 

\bibitem{papiro15a}
L. Wolfenstein, Phys. Rev. Lett. {\bf 51} (1983) 1945

\bibitem{papiro15}
G. Barenboim, F. J. Botella and O. Vives, Nucl. Phys. {\bf B613} (2001) 285 

\bibitem{papiro16}
F. del Aguila, J. A. Aguilar-Saavedra and G. C. Branco, Nucl. Phys. {\bf B510}
(1998) 39 

\bibitem{papiro22}
D. Atwood, S. Bar-Shalom, G. Eilam and A. Soni, Phys. Rept. {\bf 347} (2001) 1

\bibitem{papiro19}
T. Han, K. Whisnant, B.-L. Young and X. Zhang, Phys. Rev. {\bf D55} (1997) 7241 

\bibitem{papiro20}
M. Hosch, K. Whisnant and B.-L. Young, Phys. Rev. {\bf D56} (1997) 5725 

\end{thebibliography}
\end{document}